\documentclass[12pt]{article}
\usepackage{graphicx}
\input epsf
\usepackage{amssymb}
\parskip=\medskipamount
\def\la{\lambda}  
\def\te{\theta}   

\def\k{\kappa}    
   
\def\IC{\relax{\rm l\kern-.50 em C}}
\def\IE{\relax{\rm l\kern-.16 em E}}
\def\IK{\relax{\rm l\kern-.18 em K}}
\def\IL{\relax{\rm I\kern-.18 em L}}
\def\IN{\relax{\rm I\kern-.18 em N}}
\def\IR{\relax{\rm I\kern-.18 em R}}
\def\ii{\rm i\,}
\font\tenfrak=eufm10  \font\sevenfrak=eufm7  \font\fivefrak=eufm5
\newfam\frakfam
\textfont\frakfam=\tenfrak
\scriptfont\frakfam=\sevenfrak
\scriptscriptfont\frakfam=\fivefrak

\newtheorem{proposicion}{Proposition}

\def\wt{\widetilde}
\def\frac#1#2{{#1\over #2}}
\def\fracpd#1#2{\frac{\partial #1}{\partial #2}}

\def\Cos{\mathop{\rm C}\nolimits}    
\def\Sin{\mathop{\rm S}\nolimits}    
\def\Tan{\mathop{\rm T}\nolimits}    
\def\arcsinh{\mathop{\rm arcsinh}\nolimits} 

\begin{document}

\title{ The quantum  free particle on spherical and hyperbolic spaces:
 A curvature dependent approach. }

\author{
Jos\'e F. Cari\~nena$\dagger\,^{a)}$,
Manuel F. Ra\~nada$\dagger\,^{b)}$,
Mariano Santander$\ddagger\,^{c)}$ \\
$\dagger$
   {\sl Departamento de F\'{\i}sica Te\'orica and IUMA, Facultad de Ciencias} \\
   {\sl Universidad de Zaragoza, 50009 Zaragoza, Spain}  \\
$\ddagger$
   {\sl Departamento de F\'{\i}sica Te\'orica, Facultad de Ciencias} \\
   {\sl Universidad de Valladolid,  47011 Valladolid, Spain}
}
\date{4 Jun 2011}

\maketitle
\begin{abstract}
The quantum free particle on the sphere $S_\kappa^2$ ($\kappa>0$) 
and on the hyperbolic plane $H_\kappa^2$ ($\kappa<0$) is studied 
using a formalism that considers the curvature $\k$ as a parameter. 
The first part is mainly concerned with the analysis of some geometric
formalisms appropriate for the description of the dynamics on the
spaces  ($S_\kappa^2$, $\IR^2$, $H_\kappa^2$) and with the  the transition
from the classical $\kappa$-dependent system to the quantum one using
the quantization of the Noether momenta. The Schr\"odinger
separability and the quantum superintegrability are also discussed.
The second part is devoted to the resolution of the $\kappa$-dependent
Schr\"odinger equation.  First the characterization of the $\kappa$-dependent `curved' plane waves is analyzed and then the specific properties of the spherical
case are studied with great detail.  It is proved that if $\kappa>0$
then a discrete spectrum is obtained. The wavefunctions, that are related with a $\kappa$-dependent family of orthogonal polynomials, are explicitly obtained.
\end{abstract}

\begin{quote}
{\sl Keywords:}{\enskip}  Quantization. Quantum mechanics on
Spaces of constant curvature. 
Plane waves on spaces of constant  curvature. Orthogonal polynomials.

{\sl Running title:}{\enskip}
The quantum free particle on spaces with curvature.

{\it PACS numbers:}
{\enskip}03.65.-w, {\enskip}03.65.Ge, {\enskip}02.30.Gp, {\enskip}02.30.Ik

{\it MSC Classification:} {\enskip}81Q05, {\enskip}81R12,
{\enskip}81U15, {\enskip}34B24
\end{quote}
{\vfill}

\footnoterule
{\noindent\small
$^{a)}${\it E-mail address:} {jfc@unizar.es}  \\
$^{b)}${\it E-mail address:} {mfran@unizar.es} \\
$^{c)}${\it E-mail address:} {msn@fta.uva.es}
\newpage

\section{Introduction }

The correct formulation of quantum mechanics on spaces of constant 
curvature is a problem that can lead to important difficulties. There are 
are some fundamental quantum questions, well stated in the Euclidean 
space, that become difficult to formulate on a curved space. 
The study of these questions is important, not only for extending our knowledge of certain fundamental points of quantum mechanics, but also because it is very convenient for the construction of more general relativistic theories \cite{Birrell},\cite{Parker}. 
In addition,  this matter has become also important for  the study of certain questions arising in  applied nonrelativistic quantum mechanics. We mention here two examples related with two dimensional quantum mechanics and with condensed matter  physcis. 
In the first case (motion of a particle on a two-dimensional surface) the existence of Landau levels for the motion of a charged particle under perpendicular magenetic fields has been also studied  for the case of non-Euclidean geometries \cite{Com87}-\cite{FaSh04}. 
Concerning the second point, the study of quantum dots has also lead to the use of models based in quantum mechanics in spaces of constant curvature \cite{Gritsev01}-\cite{StoTu10}.

The first step was probably given by Schr\"odinger who made use of a 
factorization method \cite{Sch40} for the study of the Hydrogen atom in 
a spherical geometry.  Then Infeld \cite{In41} and  Stevenson \cite{St41} 
studied the same system  and Infeld and  Schild \cite{InSc45} considered 
this problem in an open universe of constant negative curvature. 
Other more recent papers on the Hydrogen atom in a  curved space 
are \cite{BaInJ87}-\cite{NiSaR99}.   Other authors  (see e.g.
\cite{BaNe03}-\cite{Gi07} and references therein) studied  the
quantum oscillator on curved spaces. We also mention that the 
path integral formulation has been also studied in curved spaces
\cite{Gr94}-\cite{LeMeBo07}.

On the other side Higgs studied in  1979, but from the point of
view of classical mechanics,  the existence of dynamical
symmetries in a spherical geometry \cite{Hi79}.  In fact his study
was mainly focussed on the existence of the spherical versions of
the Runge-Lenz vector (Kepler) and the Fradkin tensor (harmonic
oscillator). Since then a certain number of authors \cite{Le79}-\cite{KudrySigma10}  have considered  these questions or some
other properties characterizing the Hamiltonian systems defined on
curved spaces. We recall that the Kepler and the harmonic
oscillator are two systems separable in several different
coordinate systems and because of this they are superintegrable
with quadratic constants of motion. It has been proved
\cite{GrPoSi95}-\cite{BaHe09JPa} the existence of other not so
simple potentials (noncentral) that are also multiply separable on
spaces with curvature.

We also mention the study of polygonal billiards (systems enclosed by arcs of geodesics) on surfaces with curvature \cite{BaVo86},\cite{SpSa99}; one of the main points is that some simple motions, that are integrable in the Euclidean case, can become ergodic when the  curvature is negative. The quantum version of these systems leads to the study of  chaoticity in quantum systems.

The present article is concerned with the study of the quantum
free particle on spherical and hyperbolic spaces. This problem is 
usually considered as rather simple in the Euclidean case mainly 
because the solutions are plane-wave states that are in fact momentum 
eigenfunctions, that is, eigenstates of the linear momentum operator. 
The plane waves are thus simultaneous eigenfunctions of energy and 
linear momentum. Nevertheless the situation is much more complicate 
in a space with curvature mainly for two reasons.  Firstly  because 
the canonical  momenta $p_i$ do not coincide with the Noether momenta $P_i$. 
Secondly because the Noether momenta do not Poisson commute 
(Classical mechanics) and the corresponding self-adjoint quantum 
versions $\widehat{P_i}$ do not commute as operators. 
We can also add to these two points that a plane-wave is an Euclidean  concept; therefore the meaning of plane-wave in a curved space  is not clear and a new  more general definition must be introduced.

The main goal of this paper is to solve the problem by obtaining all the results making use of a curvature dependent approach. In fact, one of the main characteristics of this paper  is that it consider the  curvature $\k$ as a parameter; that is, it presents all the mathematical expressions in a $\k$-dependent way. In fact, this $\k$-dependent approach was already used in some previous related classical \cite{RaSa02b,RaSa03,CRSS04NonLin} (see also \cite{DoZi91}) and quantum studies \cite{CRS07AnPh1}-\cite{CRS07Jmp}. 

We begin the paper with  the analysis of some $\k$-dependent 
geometric  formalisms appropriate for the description of the dynamics 
on the spaces ($S_\k^2$, $\IR^2$, $H_\k^2$) with constant curvature 
$\k$ and this is done according with the study carried on in Ref.  \cite{CRS07AnPh2}. 
The first sections  present a joint approach to both  spherical ($\k>0$) and
hyperbolic dynamics  ($\k<0$) in such a way that the standard
Euclidean dynamics  just appears as the particular $\k=0$ case.
Then, the more specific properties are studied with detail but in
separate sections. 
After the first introductory paragraphs,  the rest of the article is mainly 
concerned with the following two points:

\begin{itemize}

\item  Transition from the classical $\k$-dependent system to the
quantum one using as an approach the quantization of the Noether
momenta.

\item  Exact resolution of the $\k$-dependent Schr\"odinger
equation, $\k$-dependent plane waves, families of new  $\k$-dependent
orthogonal polynomials, and existence of bound states.

\end{itemize}

It is interesting to remark that the curvature $\k$ introduce some 
new coefficients in the kinetic  term so that the problem of quantizing 
a system defined in a space with constant curvature can be related with  
the problem  of quantizing a system with a position-dependent mass.

   In more detail, the plan of the article is as follows:
In Sec. 2 we study the classical system, the quantization and the
separability of the Schr\"odinger equation. In Sec. 3 we discuss
the existence of another geometric description. Sec. 4 is devoted
to the spherical $\k>0$ case and Sec. 5 to the analysis of the
eigenfunctions $\Psi_{m,n}$ and energies $E_{m,n}$. The hyperbolic
$\k<0$ case is studied in Sec. 6.  In Sec. 7 we briefly analyze the
existence of an alternative approach and its relation with the
presence of the angular momentum. Finally, in Sec. 8 we make some
final comments.

\section{Spaces of constant curvature, $\k$-dependent formalism and
quantization } \label{Sec2}

  In what follows, all the mathematical expressions will depend of
the curvature $\k$ as a parameter, in such a way that for $\k>0$,
$\k=0$, or $\k<0$, we will obtain the corresponding property particularized
for the dynamical system on the sphere, on the Euclidean plane,
or on the hyperbolic plane respectively.

The relations among several different possible approaches to the
Lagrangian or Hamiltonian dynamics on spaces with curvature are
discussed with a certain detail in \cite{CRS07AnPh2,CRS07Jmp};
next, we summarize in the following points the relation between
the approach presented in this paper and the Higgs approach.
First, we recall that the three spaces with constant curvature,
sphere $S_{\k}^2$ ($\k>0$), Euclidean plane $\IE^2$, and
hyperbolic plane $H_{\k}^2$ ($\k<0$), can be considered as three
different situations inside a family of Riemannian manifolds
$M_{\k}^2=(S_{\k}^2,\IE^2,H_{\k}^2)$ with the curvature
$\k\in\IR$ as a parameter (it seems that this geometric idea was
first introduced by Weierstrass and Killing \cite{DoZi91}). In
fact, if we make use of the following $\kappa$-dependent
trigonometric (hyperbolic) functions
$$
 \Cos_{\k}(x) = \cases{
  \cos{\sqrt{\k}\,x}       &if $\k>0$, \cr
  {\quad}  1               &if $\k=0$, \cr
  \cosh\!{\sqrt{-\k}\,x}   &if $\k<0$, \cr}{\qquad}
  \Sin_{\k}(x) = \cases{
  \frac{1}{\sqrt{\k}} \sin{\sqrt{\k}\,x}     &if $\k>0$, \cr
  {\quad}   x                                &if $\k=0$, \cr
  \frac{1}{\sqrt{-\k}}\sinh\!{\sqrt{-\k}\,x} &if $\k<0$, \cr}
$$
and
$$
  \Tan_\k(x) = \frac{ \Sin_{\k}(x)}{ \Cos_{\k}(x)} \,,
$$
then the expression of the differential arc length element in
geodesic polar coordinates $(\rho,\phi)$ on  $M_{\k}^2$ can be written
as follows
$$
 ds_{\k}^2 = d\rho^2 + \Sin_\k^2(\rho)\,d{\phi}^2 \,,
$$
so it reduces to
$$
 ds_1^2 =    d\rho^2 + (\sin^2 \rho)\,d{\phi}^2 \,,{\quad}
 ds_0^2 =    d\rho^2 + \rho^2\,d{\phi}^2 \,,{\quad}
 ds_{-1}^2 = d\rho^2 + (\sinh^2\rho)\,d{\phi}^2\,,
$$
in the three particular cases of the unit sphere, the Euclidean
plane, and the `unit` Lobachewski plane (Note that $\rho$ denotes the
distance along a geodesic on the manifold $M_{\k}^2$ and not the
radius of a sphere). If we make use of this formalism then the
Lagrangian of the geodesic motion (free particle) on $M_{\k}^2$ is given by
\cite{RaSa02b,RaSa03,CRS07Jmp}
\begin{equation}
 \IL(\k) = (\frac{1}{2})\left(v_\rho^2 + \Sin_\k^2(\rho) v_\phi^2\right)
\end{equation}

\begin{enumerate}

\item{} If we consider the $\k$-dependent change $\rho \to\,r=\Sin_\k(\rho)$
then the Lagrangian $\IL(\k)$ becomes
$$
 L(\k) = \frac{1}{2}\,\Bigl(\frac{v_r^2}{1 - \k\,r^2} + r^2v_\phi^2 \Bigr)
$$
and, if we change to Cartesian coordinates, we arrive to
$$
 L(\k)  = \frac{1}{2}\,\Bigl(\frac{1}{1 - \k\,r^2} \Bigr)
 \Bigl[\,v_x^2 + v_y^2 - \k\,(x v_y - y v_x)^2 \,\Bigr]  \,,{\quad}
 r^2 = x^2+y^2\,.
$$
This function is just the Lagrangian studied in Ref.
\cite{CRSS04NonLin} at the classical level and in
\cite{CRS07AnPh1}-\cite{CRS07Jmp} at the quantum level (it can
also be obtained as the two-dimensional version of the kinetic
term of the one-dimensional Lagrangian $L(x,v_x;\la)$ for the
nonlinear equation of Mathews and Lakshmanan
\cite{MaLa74,LaRa03}).

\item{} If we consider the $\k$-dependent change $\rho \to\, r' =
\Tan_\k(\rho)$ then the Lagrangian $\IL(\k)$ becomes
$$
 L_H(\k) = \frac{1}{2}\,\Bigl(\frac{v_r'^2}{(1 + \k\,r'^2)^2} +
 \frac{r'^2 v_\phi^2}{(1 + \k\,r'^2)} \,\Bigr)  \,,
$$
and, if we change to Cartesian coordinates, we arrive to
$$
 L_H(\k)  = \frac{1}{2}\, \frac{1}{(1 + \k \,r'^2)^2}
 \Bigl[\,v_x^2 + v_y^2 + \k \,(x v_y - y v_x)^2 \,\Bigr]  \,,{\quad}
 r'^2 = x^2+y^2\,,
$$
that coincides with the kinetic term of the Lagrangian introduced
by Higgs in Ref. \cite{Hi79} and studied later on by other authors
\cite{Le79}-\cite{BoDaKo94} (the study of Higgs was
originally limited  to a spherical geometry but the idea can also
be applied to the hyperbolic space).
\end{enumerate}

 Hence the three Lagrangians, $\IL(\k)$, $L(\k)$ and $L_H(\k)$,
are related by diffeomorphisms, must be considered as dynamically
equivalent, and they can be alternatively used as a starting point
for the construction of the Hamiltonian quantum system. In the
following we will make use of the  Hamiltonian dynamics determined
by the $\k$-dependent Lagrangian denoted by $L(\k)$  with
coordinates $(x,y)$.

At this point we make the following observation. It is frequent to present the geometric approach to the hyperbolic plane in two steps: (i) First, consider a two dimensional pseudosphere (the upper sheet of a two-sheeted hyperboloid of revolution) inside a three-dimensional  space with Minkowskian metric. 
(ii) Then the two-dimensional model of the hyperbolic space is obtained by projection on the two-dimensional plane. The approach presented in this paper is more direct and intrisic (in differential geometric terms) and it presents directly the hyperbolic space as the manifold $M_{\k}^2=(S_{\k}^2,\IE^2,H_{\k}^2)$ endowed with the appropriate $\kappa$-dependent metric.

\subsection{Killing vectors,  Noether symmetries and Noether momenta }

We start with the following expression for the differential element
of distance  on the family $M_{\k}^2=(S_{\k}^2,\IE^2,H_{\k}^2)$
\begin{equation}
  ds_{\k}^2 = \Bigl(\frac{1}{1 - \k\,r^2}\Bigr)
 \Bigl[\,(1 - \k\,y^2)\,dx^2 + (1 - \k\,x^2)\,dy^2 + 2 \k\,x y \,dx
\,dy\,\Bigr] \,,\quad  r^2 = x^2+y^2\,,  \label{dsk(xy)}
\end{equation}
that can also be written as
\begin{equation}
 ds_{\k}^2 = \Bigl(\frac{1}{1 - \k\,r^2}\Bigr)
 \Bigl[\,dx^2 + dy^2 - \k\,(x\, dy - y\, dx)^2\,\Bigr]\,.
\end{equation}
Then the following three vector fields
$$
  X_1(\k) = \sqrt{\,1-\k\,r^2\,}\,\,\fracpd{}{x}  \,,\qquad
  X_2(\k) = \sqrt{\,1-\k\,r^2\,}\,\,\fracpd{}{y}  \,,\qquad
  X_J   = x\,\fracpd{}{y} - y\,\fracpd{}{x}\,,
$$
are  Killing vector fields, that is, infinitesimal generators of
isometries of the $\k$-dependent metric $ds_{\k}^2$. The Lie
brackets of these vector fields are given by
$$
 [X_1(\k)\,,X_2(\k)] =  \k\,X_J \,,{\quad}
 [X_1(\k)\,,X_J] =   -\,X_2 \,,{\quad}
 [X_2(\k)\,,X_J] =  X_1 \,,
$$
so that they close, depending of the sign of $\k$, the Lie algebra
of the group of isometries of
the Spherical, Euclidean and Hyperbolic spaces. Notice that only
when $\k=0$ (Euclidean space) $X_1$ and $X_2$ commute.

The geodesic motion on $M_{\k}^2=(S_{\k}^2,\IE^2,H_{\k}^2)$ is 
determined by a Lagrangian $L$ reduced to the
$\k$-dependent kinetic term $T(\k)$ without any potential
\begin{equation}
 L =  T(\k) = \frac{1}{2}\,\Bigl(\frac{1}{1 - \k\,r^2} \Bigr)
 \Bigl[\,v_x^2 + v_y^2 - \k\,(x v_y - y v_x)^2 \,\Bigr]  \,,\quad
 r^2 = x^2+y^2\,,
\end{equation}
where the parameter $\k$ can take both positive and negative values;
of course it is clear that in the spherical case  as $\k>0$,  the
function (and the associated dynamics) will have a singularity at
$1 - \k\,r^2=0$; so in this case  we shall restrict the study of
the dynamics to the region $r^2<1/\k$ where
the kinetic energy function is positive definite.
This free-particle Lagrangian  is invariant under the action of the three 
vector fields $X_1(\k)$, $X_2(\k)$, and $X_J$, in the sense that,
if we denote by $X_r^t$, $r=1,2,J$, the natural lift to the tangent
bundle (phase space $\IR^2{\times}\IR^2$) of the vector field $X_r$,
\begin{eqnarray*}
  X_1^t(\k) &=& \sqrt{\,1-\k\,r^2\,}\,\,\fracpd{}{x}
  -  \k\,\Bigl(\frac{x v_x + y v_y}{\sqrt{1-\k\,r^2\,}}\Bigr)\fracpd{}{v_x} \,,\cr
  X_2^t(\k) &=& \sqrt{\,1-\k\,r^2\,}\,\,\fracpd{}{y}
  -  \k\,\Bigl(\frac{x v_x + y v_y}{\sqrt{1-\k\,r^2\,}}\Bigr)\fracpd{}{v_y} \,,\cr
  X_J^t &=& x\,\fracpd{}{y} - y\,\fracpd{}{x}
  +  v_x\,\fracpd{}{v_y} - v_y\,\fracpd{}{v_x} \,,
\end{eqnarray*}
then the Lie derivatives of $T(\k)$ with respect to $X_r^t(\k)$
vanish, that is
$$
 X_r^t(\k)\Bigl(T(\k)\Bigr) = 0   \,,\quad  r=1,2,J.
$$
They represent three exact Noether symmetries for the geodesic
motion. If we denote by $\te_L$ the Lagrangian one-form
$$
  \te_L  =  \Bigl(\frac{v_x + \k\,J y}{1 - \k\,r^2}\Bigr)\,dx
        + \Bigl(\frac{v_y - \k\,J x}{1 - \k\,r^2}\Bigr)\,dy \,, 
$$
then the associated Noether constants of the motion $ P_1(\k)$, $P_2(\k)$
and $J$, are given
by
\begin{eqnarray*}
 P_1(\k) &=& i\bigl(X_1^t(\k)\bigr)\,\theta_L
          = \frac{v_x + \k\,J y}{\sqrt{\,1 - \k\,r^2\,}}  \cr
 P_2(\k) &=& i\bigl(X_2^t(\k)\bigr)\,\theta_L
          = \frac{v_y - \k\,J x}{\sqrt{\,1 -\k\,r^2\,}}  \cr
 J &=& i(X_J^t)\,\theta_L = x v_y - y v_x
\end{eqnarray*}

\subsection{$\k$-dependent Hamiltonian and Quantization  }

 The Legendre transformation 
leads to the following expression for the the $\k$-dependent Hamiltonian  
\begin{equation}
 H(\k)  =   \bigl(\frac{1}{2}\bigr)\,\Bigl[\,p_x^2 + p_y^2
 - \k\,(x p_x + y p_y)^2 \Bigr] \,.   \label{Hkcl(xy)}
\end{equation}
The three Noether momenta become
$$
 P_1(\k) = \sqrt{\,1 - \k\,r^2\,} \,p_x \,,\quad
 P_2(\k) = \sqrt{\,1 - \k\,r^2\,} \,p_y \,,\qquad
 J = x p_y - y p_x\,,
$$
with Poisson brackets
$$
 \{P_1(\k)\,,P_2(\k)\} = \k\,J  \,,{\quad}
 \{P_1(\k)\,,J\} = -  P_2(\k) \,,{\quad}
 \{P_2(\k)\,,J\} =  P_1(\k) \,,
$$
and such that 
$$
 \{P_1(\k)\,,H(\k)\} = 0  \,,{\quad}
 \{P_2(\k)\,,H(\k)\} = 0 \,,{\quad}
 \{J\,,H(\k)\} = 0 \,.
$$
A very important property is that the Hamiltonian can be written as
a function of the three Noether momenta. It is given by
\begin{equation}
 H(\k)  =  (\frac{1}{2})\,\bigl[\,P_1^2 + P_2^2 + \k\,J^2 \,\bigr] \,. 
 \label{Hkcl(PPJ)}
\end{equation}
We note that $H(\k)$ is just the Casimir of the above Poisson algebra.

The following task is to find the appropriate quantum mechanical Hilbert space
and this means to obtain a measure that reduces to the standard one when $\k\to 0$. 
An important property is that the only  measure on the space $\IR^2$, with coordinates $(x,y)$, that is invariant under the action of the three vector fields
$X_1(\k)$, $X_2(\k)$, and $X_J$, is given by
$$
  d\mu_\k = \Bigl(\frac{1}{\sqrt{1-\k\,r^2}}\Bigr)\,dx\,dy \,,
$$
up to a constant factor (see Ref.  \cite{CRS07AnPh2} for a proof).

This property means that the quantum Hamiltonian  must be 
self-adjoint, not in the standard space $L^2(\IR^2)$, but in the Hilbert space $L_\k^2(d\mu_\k)$ defined as 
\begin{itemize}
\item[(i)]  In the hyperbolic $\k<0$ case, the space $L_\k^2(d\mu_\k)$ is $L^2(\IR^2,d\mu_\k)$. 
\item[(ii)] In the spherical $\k>0$ case, the space $L_\k^2(d\mu_\k)$ is $L_0^2(\IR_\k^2,d\mu_\k)$ where $\IR_\k^2$ denotes the region $r^2\le 1/\k$ and the subscript means that the functions must vanish at the  boundary of this region 
\end{itemize}
(the question of the boundary conditions will be discussed with more detail when considering the Sturm-Liouville problem). 
For obtaining the expression  of the operator 
$\widehat{H}(\k)$ we first consider  the operators $\widehat{P_1}$
an $\widehat{P_2}$, representing the quantum version of of the
Noether momenta momenta $P_1$ an $P_2$, that must be also self-adjoint 
 in the space $L_\k^2(d\mu_\k)$. They are given by
\begin{eqnarray}
 \widehat{P_1} &=& -\,i\,\hbar\,\sqrt{1 - \k\,r^2}\,\fracpd{}{x} \,,\cr
 \widehat{P_2} &=& -\,i\,\hbar\,\sqrt{1 - \k\,r^2}\,\fracpd{}{y} \,.
\nonumber \end{eqnarray} 
Then we arrive to the following correspondence
\begin{eqnarray*}
 P_1^2 \ &\to&\ -\,\hbar^2\,
 \Bigl(\sqrt{1 - \k\,r^2}\,\fracpd{}{x}\Bigr)
 \Bigl(\sqrt{1 - \k\,r^2}\,\fracpd{}{x}\Bigr) \,,\cr
 P_2^2 \ &\to&\ -\,\hbar^2\,
 \Bigl(\sqrt{1 - \k\,r^2}\,\fracpd{}{y}\Bigr)
 \Bigl(\sqrt{1 - \k\,r^2}\,\fracpd{}{y}\Bigr) \,,
\end{eqnarray*}
 in such a way that the quantum  Hamiltonian operator $\widehat{H}(\k)$
$$
 \widehat{H}(\k) = (\frac{1}{2})\,\bigl[\,\widehat{P_1}^2 + \widehat{P_2}^2 + \k\,\widehat{J}^2 \,\bigr]
$$
is given by
\begin{eqnarray}
  \widehat{H}(\k) =  &-& \frac{\hbar^2}{2 m}\, \Bigl[\,
 (1 - \k\,r^2)\,\fracpd{^2}{x^2} - \k\,x\,\fracpd{}{x} \,\Bigr]
 - \frac{\hbar^2}{2 m}\, \Bigl[\,
 (1 - \k\,r^2)\,\fracpd{^2}{y^2} - \k\,y\,\fracpd{}{y} \,\Bigr]\cr
 &-& \k\, \frac{\hbar^2}{2 m}\, \Bigl[\,
 x^2\,\fracpd{^2}{y^2} + y^2\,\fracpd{^2}{x^2} - 2 x y \,\fracpd{^2}{x \,\partial y} - x\,\fracpd{}{x} - y \,\fracpd{}{y}\,\Bigr]   \label{Hkq(xy)}
 \end{eqnarray}

The first important property of this Hamiltonian is that it
admits the following decomposition
$$
 \widehat{H}(\k) = \widehat{H_1} + \widehat{H_2} + \k\, \widehat{J}^2 \,, 
$$
where the three partial operators $\widehat{H_1}$, $\widehat{H_2}$
y $\widehat{J}^2$ are respectively given by
\begin{eqnarray}
 \widehat{H_1}(\k) &=& - \frac{\hbar^2}{2 m}\, \Bigl[\,
 (1 - \k\,r^2)\,\fracpd{^2}{x^2} - \k\,x\,\fracpd{}{x} \,\Bigr]    \cr
 \widehat{H_2}(\k) &=&  - \frac{\hbar^2}{2 m}\, \Bigl[\,
 (1 - \k\,r^2)\,\fracpd{^2}{y^2} - \k\,y\,\fracpd{}{y} \,\Bigr]   \cr
\widehat{J}^2 &=& -\, \frac{\hbar^2}{2 m}\, \Bigl[\,
 x^2\,\fracpd{^2}{y^2} + y^2\,\fracpd{^2}{x^2} - 2 x y \,\fracpd{^2}{x \,\partial y} - x\,\fracpd{}{x} - y \,\fracpd{}{y}\,\Bigr]
\nonumber \end{eqnarray}
in such a way that the total Hamiltonian $\widehat{H}$ commutes,
for any value of the parameter $\k$,  with each one
of the three partial terms
$$
 \bigl[ \widehat{H}(\k)\,,\widehat{H_1}(\k) \bigr] = 0\,,{\hskip 10pt}
 \bigl[\widehat{H}(\k)\,,\widehat{H_2}(\k) \bigr] = 0\,,{\hskip 10pt}
 \bigl[ \widehat{H}(\k)\,,\widehat{J}^2 \bigr] = 0\,.
$$
The vanishing of these three commutators means that the
$\k$-dependent  Hamiltonian (\ref{Hkq(xy)}) describes a quantum
superintegrable system \cite{LeVi95}-\cite{CaNedO06}. This
property was well known in the Euclidean $\k=0$ case but now it
appears in a different form because of presence of the term $\k\,
\widehat{J}^2$.

\subsection{Schr\"odinger equation and Separability}

 Now, if we consider the Sch\"rodinger equation
\begin{equation}
 \widehat{H}\,\Psi = E\,\Psi\,,   \label{EcSchxy}
\end{equation}
as we have the following property
$$
 \bigl[ \widehat{H}_1\,,\widehat{H}_2 + \k\, \widehat{J}^2 \bigr] = 0\,,{\hskip 10pt}
 \bigl[ \widehat{H}_1 + \k\, \widehat{J}^2\,,\widehat{H}_2 \bigr] = 0\,,{\hskip 10pt}
 \bigl[ \widehat{H}_1 + \widehat{H}_2\,, \widehat{J}^2 \bigr] =  0\,,
$$
then, we have three different sets of compatible observables and therefore
three different ways of obtaining a Hilbert basis of common eigenstates.
\begin{enumerate}
\item{} The two operators $\widehat{H_1}$ and $\widehat{H_2} +
\k\, \widehat{J}^2$ are a (complete) set of commuting observables;
therefore they represent two quantities that can be simultaneously
measured. Thus, the first way of looking for $\Psi$ is as a
solution of the following two equations
$$
\widehat{H_1}\,\Psi = e_1\,\Psi  \,,{\hskip 10pt}
\bigl(\widehat{H_2} + \k\, \widehat{J}^2\bigr)\,\Psi = e_{2j}\,\Psi  \,.
$$
In this case the total energy is given by $E = e_1 + e_{2j}$ and
the associated wave function can be denoted by $\Psi(e_1,
e_{2j})$.

\item{} The two operators $\widehat{H_1} + \k\, \widehat{J}^2$ and
$\widehat{H_2}$ are a (complete) set of commuting observables. Thus,
the second way of looking for $\Psi$ is as a solution of the
following two equations
$$
\bigl(\widehat{H_1} + \k\, \widehat{J}^2\bigr)\,\Psi =
e_{1j}\,\Psi  \,,{\hskip 10pt} \widehat{H_2}\,\Psi = e_2\,\Psi \,.
$$
In this case we have $E = e_{1j} + e_2$ and $\Psi$ can be denoted
by $\Psi(e_{1j}, e_2)$.

\item{} The third (complete) set of commuting observables is
provided by $\widehat{H_1} + \widehat{H_2}$ and $\widehat{J}^2$.
So in this case we have
$$
 \bigl(\widehat{H_1} + \widehat{H_2}\bigr)\,\Psi = e_{12}\,\Psi  \,,{\hskip 10pt}
 \widehat{J}^2\,\Psi = e_j\,\Psi   \,.
$$
Thus, the two physically measurable quantities are $e_{12}$ and
the angular momentum $j$, the total energy is given by $E = e_{12}
+ \k\,e_j$ and the wave function so defined can be denoted by
$\Psi(e_{12},e_j)$.
\end{enumerate}

The existence of these three alternative descriptions arises
from the presence of the term $\k\,\widehat{J}^2$ inside the
kinetic part of the Hamiltonian. Notice that the second approach
can be considered as symmetric to the first one. Nevertheless, although they
are closely related, they lead however to different solutions with
different properties; that is, $\Psi(e_1, e_{2j}) \ne \Psi(e_{1j},
e_2)$. This fact is a consequence of the nonlinear character of
the model since in the linear limit, when $\k\to 0$, then both
descriptions coincide.

The  $\k$-dependent metric $ds_\k^2$ is not diagonal in the
coordinates $(x,y)$ and the Schr\"odinger equation (\ref{EcSchxy})
is not separable in these coordinates because of the
$\k$-dependent term. Nevertheless, the classical Hamilton-Jacobi
equation
$$
  \Bigl(\fracpd{S}{x}\Bigr)^2 + \Bigl(\fracpd{S}{y}\Bigr)^2
 - \k\,\Bigl(x\,\fracpd{S}{x} + y\,\fracpd{S}{y}\Bigr)^2 = 0
$$
and the quantum Schr\"odinger equation admit separability in the
following three different orthogonal  coordinate systems:
\begin{enumerate}
\item{} $\k$-dependent coordinates $(z_x,y)$ with $z_x$ defined by
$z_x = x/\sqrt{1 - \k\,y^2}$\,.

\item{} $\k$-dependent coordinates $(x,z_y)$ with $z_y$ defined by
$z_y = y/\sqrt{1 - \k\,x^2}$\,.

\item{} Polar coordinates $(r,\phi)$.
\end{enumerate}
At this point we recall that the existence of multiple
separability is a property directly related with
superintegrability (in fact with quadratic superintegrability).

Next we start our study with the first coordinate system. First
note that the change $(x,y)\to(z_x,y)$ transforms, in the $k>0$
spherical case,  the circular domain $x^2+y^2<1/\k$ into the
square region $z_x^2<1/\k$, $y^2<1/\k$, in the $(z_x,y)$ plane.

Using $(z_x,y)$ coordinates the two partial Hamiltonians become:

\begin{enumerate}
 \item[(i)]   The Hamiltonian  $\widehat{H}_1$ is given by
$$
   \widehat{H}_1 = -\,\frac{\hbar^2}{2 m}\, \wt{H}_1 \,,{\quad}
   \wt{H}_1 = \bigl(1 - \k\,z_x^2\bigr)\fracpd{^2}{z_x^2}
 - \bigl(\k\,z_x\bigr)\,\fracpd{}{z_x}    \,.
$$
 \item[(ii)]   The Hamiltonian  $\widehat{H}_2+ \k\, \widehat{J}^2$ given by
$$
  \widehat{H}_2+ \k\, \widehat{J}^2 = -\,\frac{\hbar^2}{2 m}\, \bigl[\wt{H}_{2} + \k\, \wt{J}^2\bigr] \,,
$$
is represented by the following differential operator
$$
   \wt{H}_{2} + \k\, \wt{J}^2= \frac{\k\,y^2}{1 - \k\,y^2}
   \Bigl[\bigl(1 - \k\,z_x^2\bigr)\fracpd{^2}{z_x^2}
 - \bigl(\k\,z_x\bigr)\,\fracpd{}{z_x}    \Bigr] +
 \Bigl[\bigl(1 - \k\,y^2\bigr)\fracpd{^2}{y^2}
 - \bigl(2\k\,y\bigr)\,\fracpd{}{y}  \Bigr]   \,, 
$$
that can be rewritten as follows
$$
   \wt{H}_{2} + \k\, \wt{J}^2= \frac{\k\,y^2}{1 - \k\,y^2}\,
   \wt{H}_1+ \Bigl[\bigl(1 - \k\,y^2\bigr)\fracpd{^2}{y^2}
 - \bigl(2\k\,y\bigr)\,\fracpd{}{y}  \Bigr]  \,. 
$$
\end{enumerate}
Consequently the $\k$-dependent Schr\"odinger equation is in fact
separable in the $(z_x,y)$ coordinates. Thus the two-dimensional
problem has been decoupled in two one-dimensional equations.
\begin{enumerate}

\item[(i)]   The Schr\"odinger equation $\widehat{H}_1\Psi =
e_1\Psi $ for the first partial Hamiltonian $\widehat{H}_1$ leads to the
following equation with derivatives with respect to the  variable
$z_x$ alone
$$
 \bigl(1 - \k\,z_x^2\bigr)\,\Psi_{z_xz_x}''   - \bigl(\k\,z_x\bigr)\,\Psi_{z_x}'  + \mu \,\Psi = 0  \,,\quad 
 \mu = \bigl(\frac{2 m }{\hbar^2}\bigr) e_1 \,. 
$$
\item[(ii)]   The Schr\"odinger equation $(\widehat{H}_2 + \k\, \widehat{J}^2)\Psi = e_{2j}\Psi $  for the second partial Hamiltonian $\widehat{H}_2 + \k\, \widehat{J}^2$ leads to the following $\mu$-dependent equation with derivatives with respect to the  variable $y$ alone
$$
-\frac{\k\,y^2}{1 - \k\,y^2}\,(\mu\Psi)  + \Bigl[\bigl(1 - \k\,y^2\bigr)\,\Psi_{yy}''  - \bigl(2\k\,y\bigr)\,\Psi_{y}'   \Bigr]+ \nu \,\Psi = 0
  \,,\quad \nu = \bigl(\frac{2 m }{\hbar^2}\bigr) e_{2j}\,.
 $$
\end{enumerate}
 Thus, if we assume that $\Psi(z_x,y)$ is a function of the form
$$
 \Psi(z_x,y)  = Z(z_x)\,Y(y) \,,
$$
then we arrive to
$$
 \bigl(1 - \k\,z_x^2\bigr)\,Z''   - \bigl(\k\,z_x\bigr)\,Z'  + \mu \,Z = 0 \,, 
$$
and
$$
  (1 - \k\,y^2)\,Y'' - \bigl( 2\k\,y\bigr)\,Y'   -  \mu\k\,
  \Bigl(\frac{y^2}{1 - \k\,y^2} \Bigr)\,Y +  \nu\,Y = 0   \,.
$$

\section{An alternative approach: Parallel  geodesic coordinates}

The study presented in Sec. \ref{Sec2} (symmetries,  quantization, separation of variables) was constructed starting with the expression (\ref{dsk(xy)}) for $ds_\k^2$; nevertheless, it  can alternatively be developed in other coordinate systems. Now we present in this section the results obtained when making use of parallel  geodesic coordinates $(u,y_\k)$ (see the Appendix for more information on the geometric
origin of this particular system of coordinates and  Ref. \cite{RaSa02b,RaSa03} (and references therein) for some papers that make use of this formalism).

  The following expression written in $(u,y_\k)$ parallel coordinates
\begin{equation}
  ds_\k^2 = \Cos_\k^2(y_\k) du^2 + dy_\k^2 \,.
\end{equation}
represents the differential element of distance on the spaces
$(S_\k^2,\IE^2,H_\k^2)$ with constant curvature $\k$.
So a standard lagrangian (kinetic term minus a potential function)
has the following form
 $$
  L(\k) = (\frac{1}{2})\bigl(\Cos_\k^2(y_\k)\,v_u^2 + v_{y_{\k}}^2\bigr)
      -  U(u,y_\k,\k) \,, 
$$
in such a way that the Euclidean system is just given by the particular
value of $L(\k)$ in $\k=0$
 $$
  \lim_{\k\to 0}\,L(\k) = (\frac{1}{2})\,(v_x^2 + v_y^2) -  V(x,y)
  \,,{\qquad} V(x,y) = U(x,y_\k,\k)\Bigl|_{\k=0} \,.
 $$
The three $\k$-dependent Killing vector, $Y_1$, $Y_2(\k)$, and $Y_J(\k)$,
have now the following expressions in parallel coordinates
\begin{eqnarray*}
   Y_1  &=&  \fracpd{}{u}\,,\cr
   Y_2(\k) &=& \k\,\Sin_\k(u)\Tan_\k(y_\k)\,\fracpd{}{u}
    + \Cos_\k(u)\,\fracpd{}{y_\k}\,, \cr
   Y_J(\k) &=&  \Cos_\k(u)\Tan_\k(y_\k)\,\fracpd{}{u}
    - \Sin_\k(u)\,\fracpd{}{y_\k}\,,
\end{eqnarray*}
($Y_1$ is now $\k$-independent) and the associated linear
constants of motion are given by
\begin{eqnarray*}
  P_1(\k) &=&  \Cos_\k^2(y_\k)\,v_u \,,   \cr
  P_2(\k) &=&  \k\,\Sin_\k(u)\Cos_\k(y_\k)\Sin_\k(y_\k)\,v_u
  + \Cos_\k(u)\,v_{y_{\k}}\,, \cr
  J(\k)   &=& \Cos_\k(u)\Cos_\k(y_\k)\Sin_\k(y_\k)\,v_u
  - \Sin_\k(u)\,v_{y_{\k}}\,.
\end{eqnarray*}

 Then, when moving to the Hamiltonian formalism we obtain that the Noether
momenta are given by
 \begin{eqnarray*}
  P_1(\k) &=&  p_u \,,   \cr
  P_2(\k) &=&  \k\,\Sin_\k(u)\Tan_\k(y_\k)\, p_u + \Cos_\k(u)\,p_{y_{\k}}\,, \cr
  J(\k)   &=& \Cos_\k(u)\Tan_\k(y_\k)\, p_u  - \Sin_\k(u)\,p_{y_{\k}}\,,
\end{eqnarray*}
in such a way tha the geodesic Hamiltonian
$$
 H(\k) = (\frac{1}{2})\Bigl(\,\frac{p_u^2}{\Cos_\k^2(y_\k)} + p_{y_{\k}}^2\,\Bigr) \,,
$$
can also be written as
$$
 H(\k) = (\frac{1}{2})\,\bigl(P_1^2 + P_2^2 + \k\,J^2 \bigr) \,.
$$

The  $\k$-dependent  measure $d\mu_\k$, invariant under the action of the three vector fields $Y_1$, $Y_2$ and $Y_J$, is given by 
$$ 
  d\mu_\k = \Cos_\k(y_\k)\,du\,dy_\k \,, 
$$  
and the transition from classical to quantum mechanics via the Noether momenta is now represented by the following correspondence
\begin{eqnarray*}
 P_1\ \to\ \widehat{P_1} &=&   -\,i\,\hbar\,\fracpd{}{u}  \,,\cr
 P_2\ \to\ \widehat{P_2} &=&   -\,i\,\hbar\,\Bigl(\k\,\Sin_\k(u)\Tan_\k(y_\k)\,\fracpd{}{u} + \Cos_\k(u)\,\fracpd{}{y_\k}\Bigr)  \,,\cr
 J\ \to\ \widehat{J} &=&   -\,i\,\hbar\,\Bigl(\Cos_\k(u)\Tan_\k(y_\k)\,\fracpd{}{u} - \Sin_\k(u)\,\fracpd{}{y_\k}\Bigr)  \,,
\end{eqnarray*}
so that we arrive to
\begin{enumerate}
\item   The quantum operator $\widehat{H}_1$ is given by
$$
   \widehat{H}_1 = -\,\frac{\hbar^2}{2 m}\, \wt{H}_1 \,,{\quad}
   \wt{H}_1 = \fracpd{^2}{u^2}  \,.
$$
\item   The quantum operator $\widehat{H}_2+ \k\, \widehat{J}^2$   given by
$$
   \widehat{H}_2+ \k\, \widehat{J}^2 = -\,\frac{\hbar^2}{2 m}\, \bigl[\wt{H}_{2} + \k\, \wt{J}^2\bigr]
$$
is represented by the following differential operator
$$
  \wt{H}_{2} + \k\, \wt{J}^2 =  \k\,\Tan_\k^2(y_\k)\,\fracpd{^2}{u^2}
  + \fracpd{^2}{y_\k^2}  - \k\,\Tan_\k(y_\k) \fracpd{}{y_\k}   \,,
$$
that can be rewritten as follows
$$
   \wt{H}_{2} + \k\, \wt{J}^2= \k\,\Tan_\k^2(y_\k)\,  \wt{H}_1
   + \Bigl[\fracpd{^2}{y_\k^2}  - \k\,\Tan_\k(y_\k) \fracpd{}{y_\k} \Bigr] \,.
$$
\end{enumerate}

In this way, the two Schr\"odinger equations for $\widehat{H}_1$
(equation for the variable $u$) and for $\widehat{H}_2+ \k\,
\widehat{J}^2$ (equation for the variable $y_\k$) are
\begin{enumerate}
 \item[(i)]   The Schr\"odinger equation $\wt{H}_1\psi = \mu\,\Psi$
 determined by the Hamiltonian $\widehat{H}_1$ leads to
$$
 \Psi_{uu}''  + \mu \,\Psi = 0  \,,{\quad}
  \mu = \bigl(\frac{2 m }{\hbar^2}\bigr) e_1\,.
$$
\item[(ii)]   The Schr\"odinger equation $(\wt{H}_{2} + \k\,
\wt{J}^2)\Psi = \nu\,\Psi$  determined by the Hamiltonian
$\widehat{H}_2+ \k\, \widehat{J}^2$ leads to
$$
-\k\,\mu\Tan_\k^2(y_\k)\,\Psi  + \Bigl[\,\Psi_{y_{\k}y_{\k}}''   - \k\,\Tan_\k(y_\k)\,\Psi_{y_\k}'   \Bigr]+ \nu \,\Psi = 0  \,,{\quad}
  \nu = \bigl(\frac{2 m }{\hbar^2}\bigr) e_{2j}\,.
$$
\end{enumerate}
 Thus, if we assume that $\Psi(u,y_\k)$ is a function of the form
$$
  \Psi(u,y_{\k})  = U(u)\,Y(y_\k) \,,
$$
then we arrive to
$$
 U_{uu}''  + \mu \,U= 0 \,,
$$
and
$$
 Y_{y_{\k}y_{\k}}'' - \k\,\Tan_\k(y_\k)\,Y_{y_\k}' + \Bigl(\nu  -\k\,\mu\Tan_\k^2(y_\k)\Bigr) \,Y= 0 \,.
$$

The $U$-equation is just the same equation that in the Euclidean
case (so the solution is a $u$-plane-wave, that is, a plane-wave
along the geodesic curve $y_\k=0$). Concerning the $Y$-equation,
it can be simplified by using the following factorization
$$
   Y = (\Cos_\k(y_\k))^{g_\k} p(y_\k) \,,{\quad} g_\k = \sqrt{\mu/\k}
   \,, {\quad} \k>0\,,
$$
so that we arrive to
$$
 p_{y_{\k}y_{\k}}'' - \k\, (1 + 2 g_\k)\,\Tan_\k(y_\k)\,p_{y_\k}'
 + \bigl(\nu -\k\,g_\k\bigr) \,p= 0 \,.
$$

\section{Spherical $\k>0$ case}

Now we return to the approach developed in Sec. \ref{Sec2} and
study the spherical $\k>0$ case.

\subsection{Resolution of the $Z$-equation}

The first equation to be solved is
\begin{equation}
 \bigl(1 - \k\,z_x^2\bigr)\,Z''   - \bigl(\k\,z_x\bigr)\,Z'  + \mu \,Z = 0 \,,
 \quad \k>0 \,.
\label{EcZe}\end{equation} This equation coincides (up to the
appropriate changes of notation) with the equation corresponding
to a one-dimensional $\k$-dependent free particle. Assuming for
$Z$ an expression of the form
$$
   Z =   e^{{\ii}u(z_x)} \,,
$$
with $u(z_x)$ a function to be determined, then we obtain that the
general solution of (\ref{EcZe}) is given by
\begin{equation}
  Z  =  A\,e^{{\ii}u(z_x)} + B\,e^{-{\ii}u(z_x)}  \,,{\quad}
  u = \frac{\sqrt{\mu}}{{\sqrt{\k}}}\,\arcsin(\sqrt{\k}\,z_x) \,,\quad
  \k>0 \,, \label{Zue}
\end{equation}
that is a well defined function for all the values of $z_x$. This
solution satisfies the appropriate Euclidean limit
$$
 \lim_{\k\to 0}  Z = A\,e^{{\ii}k_x x} + B\,e^{-{\ii}k_x x}  \,,{\quad}
 k_x = \sqrt{\mu} \,,
$$
and therefore it can be considered as representing a
$\k$-dependent curved plane-wave or a $\k$-dependent deformation
of the Euclidean plane-wave solution.

\subsection{Resolution of the $Y$-equation}

The second equation to be solved is
\begin{equation}
   (1 - \k\,y^2)\,Y'' - \bigl( 2\k\,y\bigr)\,Y'   -  \mu\k\,
   \Bigl(\frac{y^2}{1 - \k\,y^2} \Bigr)\,Y +  \nu\,Y = 0   \,,\quad \k>0 \,, 
\label{EcYe1}\end{equation} 
that, although it has certain similarity with the Eq. (\ref{EcZe}), 
it does not coincide with it (two differences: the factor $2$ in 
the coefficient of $Y'$ and the rational $\mu$-dependent term). 
The main reason for this asymmetry is that, when introducing 
separability in the Schr\"odinger equation, the angular momentum 
term $\widehat{J}^2$ was displaced into this second equation.

It can be verified that the function $\Psi_Y$ defined by
$$
  \Psi_Y = (1 - \k\,y^2)^{(1/2)g_\k}    \,,{\quad}
   g_\k = \sqrt{\frac{\mu}{\k}}  \,,
$$
satisfies the following property
$$
  \Bigl[(1 - \k\,y^2) \frac{d^2}{dy^2} - \bigl( 2\k\,y\bigr) \frac{d}{dy}   -  \mu\k\,
  \Bigl( \frac{y^2 }{1 - \k\,y^2} \Bigr)\Bigr] \Psi_Y  = - \k g_\k  \Psi_Y  \,.
$$
Thus $\Psi_Y$ represents the exact solution in the very particular
case of $\nu= \k g_\k$. This property suggests the following
factorization for the function $Y(y)$
$$
 Y(y) = p(y)\,(1 - \k\,y^2)^{(1/2)g_\k} \,,{\quad}
 g_\k = \sqrt{\frac{\mu}{\k}}\,,
$$
where the factor on the right satisfies the following Euclidean limit
$$
 \lim_{\k\to 0}(1 - \k\,y^2)^{(1/2)g_\k}
 = \exp\left[\lim_{\k\to 0}\frac{\sqrt{\mu}}{2} \,\frac{\log(1-\k y^2)}{\sqrt{\k}}\right]
 = \exp\left[\frac{\sqrt{\mu}}{2} \,\lim_{\k\to 0}\frac{-y^2}{(1-\k y^2)}(2\sqrt{\k})\right] = 1 \,.
$$
Then the equation becomes
\begin{equation}
 (1 - \k\,y^2)  p'' - 2 \k\, (1+g_\k) y \,p' + (\nu - \k g_\k) p = 0  \,,
 \quad p=p(y) \,.  \label{Ecp(y)e1}
\end{equation}

This equation is an equation of hypergeometric type and it can be
reduced to the canonical form of a hypergeometric Gauss equation
with singular points in $w=0$ and $w=1$
$$
 w (1-w)p_{ww}'' + (\la_a  + \la_b w)  \,p_w' + \la_c p = 0 \,,
$$
by making use of the change  $y\to w$ given by
$$
 w = \frac{1}{2}(1 + \sqrt{k}\, y) \,.
$$
Nevertheless, as $y=0$ is an ordinary point, it can be also  directly solved by assuming a power
expansion for the solution
$$
 p(y,\k) = \sum_{n=0}^\infty\,p_n(\k)\,y^n
  = p_0(\k) + p_1(\k)\,y + p_2(\k)\,y^2 + \dots
$$
that leads to the following $\k$-dependent recursion relation
$$
 p_{n+2}  =   \biggl[\,\frac{\k\,n(n-1) + 2\k (1+ g_\k) n -
 (\nu - \k g_\k)}{(n+2)(n+1)}\,\biggr] \ p_n \,.
$$
Note that this relation shows that, as in the particular $\k=0$
case, even power coefficients are related among themselves and the
same is true for odd power coefficients. In both cases, having in
mind that
$$
\lim{}_{n\to\infty}\,\biggl|\frac{p_{n+2}y^{n+2}}{p_ny^n}\biggr|
  = \lim\nolimits_{n\to \infty}\,\biggl|\,
  \frac{\k\,n(n-1) + 2\k (1+ g_\k) n - (\nu - \k g_\k)}{(n+2)(n+1)}\,\biggr|\,\bigl|\,y^2\bigr| =
 |\,\k\,|\,\bigl|\,y^2\bigr| \,,
$$
the radius of convergence $R$ is given by $R=1/{\sqrt{\,|\,\k\,|\,}}$. 
 Hence, when we consider the limit $\k\to0$, we recover the radius
$R=\infty$ of the Euclidean equation.

The general solution is given by a linear combination
$$
   Y(\k) = \bigl(C\,Y_{ev}(y) + D\,Y_{od}(y)\bigr)\,(1 - \k\,y^2)^{(1/2)g_\k} \,,
$$
where $Y_{ev}(y)$ is an even function and $Y_{od}(y)$ is an odd
function with $Y_{ev}(0)=1$ and $Y'_{od}(0)=1$. In the Euclidean
limit, if $c_n$ denotes  $c_n=\lim_{\k\to 0} p_n(\k,\mu,\nu)$, then the
recursion relation reduces to
$$
 c_{n+2}  =    \frac{ (-\,\nu)\,c_n}{(n+2)(n+1)}   \,,  
$$
and the solution becomes
$$
   \lim_{\k\to 0}Y(\k) = C\,\cos(\sqrt{\nu}\,y) + D\,\sin(\sqrt{\nu}\,y) \,,
$$
that can also be written as
$$
 \lim_{\k\to 0}Y(\k) = \wt{C}\,e^{{\ii}k_y y} + \wt{D}\,e^{-{\ii}k_y y}  \,,{\quad}
 k_y = \sqrt{\nu} \,.
$$
and represents  Euclidean plane-waves.
 
In the very particular case of an integer $n$ such that 
$$
 \nu - \k g_\k = \k\,n(n-1) + 2\k (1+ g_\k) n
$$
then we have $c_n  \ne  0$, $c_{n+2} = 0$, and one of the two
solutions (even or odd) becomes a polynomial of order $n$.
The  coefficient $\nu$ be given by $\nu=\nu_{n}$ with
$$
 \nu_{n} = \k[n(n+1) + g_\k(2 n + 1)]   {\qquad}
 (n\ {\rm \ is\ an\ integer\ number)}.
$$
The polynomial solutions are given by
\begin{itemize}
 \item Even index (even power polynomials):
The expressions of the first solutions ${\cal P}_j(y)$, in the
particular cases of $j=0,2,4$, are given by:
\begin{eqnarray}
 {\cal P}_0 &=&  1 \,,\cr
 {\cal P}_2 &=&  1 - \k (3 + 2 g_\k) y^2  \,,\cr
 {\cal P}_4 &=&  1 - 2 \k (5 + 2 g_\k) y^2 +  (\frac{\k^2}{3}) (5 + 2 g_\k) (7 + 2 g_\k) y^4
 \end{eqnarray}

 \item Odd index  (odd power polynomials):
The expressions of the second solutions  ${\cal P}_j(y) $,
for $j=1,3,5$, are given by:
\begin{eqnarray}
 {\cal P}_1 &=&  y  \,,\cr
 {\cal P}_3 &=&  y - (\frac{\k}{3})(5 + 2 g_\k)  y^3  \,,\cr
 {\cal P}_5 &=&   y - (\frac{2\k}{3})(7 + 2 g_\k)  y^3
 + (\frac{\k^2}{15})(7 + 2 g_\k)  (9
 + 2 g_\k)  y^5   \,.
\end{eqnarray}
\end{itemize}

\section{Wavefunctions and  eigenvalues}   \label{Sec. 5}

Let us start pointing out that 
the measure $d\mu_\k$ can be written as follows
\begin{equation}
  d\mu_\k = \Bigl(\frac{1}{\sqrt{1-\k\,r^2}}\Bigr)\,dx\,dy
  = \Bigl(\frac{dz_x}{\sqrt{1-\k\,z_x^2}}\Bigr)\,dy\,.\label{dmu2}
\end{equation}
Thus, the coordinates $(z_x,y)$ also factorize the $\k$-dependent
measure.

\subsection{Sturm-Liouville problem for the $Z$-equation}

The $\k$-dependent differential equation
$$
 a_0 Z''  + a_1\,Z'  +  a_2 Z = 0 \,,{\quad}
$$
with
$$
 a_0 =  1 - \k\,z_x^2  \,,{\quad}  a_1 = - \,\k\,z   \,,{\quad}
 a_2 = \mu \,,
$$
is not self-adjoint since $a'_0 \ne a_1$ but it can be reduced to
self-adjoint form by making use of the following integrating
factor
$$
 \mu(z_x) = \bigl(\frac{1}{a_0}\bigr)\,e^{{\int}(a_1/a_0)\,dz_x}
 = \sqrt{1 - \k\,z_x^2}  \,,
$$
in such a way that if we denote by $q=q(z_x,\k)$ and $r=r(z_x,\k)$
the following functions
\begin{eqnarray*}
   q(z_x,\k) &=& e^{{\int}(a_1/a_0)\,dy} = \sqrt{1 - \k\,z_x^2}\,,\cr
   r(z_x,\k) &=& \bigl(\frac{a_2}{a_0}\bigr)\,e^{{\int}(a_1/a_0)\,dx}
   = \frac{\mu}{\sqrt{1 - \k\,z_x^2}}   \,.
\end{eqnarray*}
then we arrive to the following  expression
\begin{equation}
 \frac{d}{dz_x}\Bigl[\,q(z_x,\k)\,\frac{dZ}{dz}\,\Bigr] + r(z_x,\k)\,Z = 0 \,.
 \label{EcSLZ(z)e}
\end{equation}
Thus, this self-adjoint equation together with appropriate
conditions for the behaviour of the solutions at the end points,
constitute a  Sturm-Liouville problem.

 If $\k$ is positive the range of the  variable $z_x$ is limited by
the restriction $z_x^2<1/\k$. In this case the problem,  defined in
the bounded interval $[-\,a_\k,a_\k]$ with $a_\k=1/\sqrt{\k}$, is
singular because the function $q(z_x,\k)$ vanishes in the two end
points $z_{x1}=-\,a_\k$ and $z_{x2}=a_\k$.  So the first condition to be
imposed  is that  the  solutions $Z(z_x,\k)$ of the problem must be
bounded functions at the two end points of the interval so that
the norm be finite. Then we note that this situation is rather
similar to the case of a particle in a one-dimensional square well
with perfectly rigid impenetrable walls at the points
$z_{x1}=-\,a_\k$ and $z_{x2}=a_\k$. 
Hence, taking into account that $u(z_{1x})=-g_\k\frac{\pi}{2}$  and 
$u(z_{2x})=g_\k\frac{\pi}{2}$, the application of the boundary
conditions at $z_{x1,2}=\pm\,a_\k$,  gives
$$  
 \wt{A} \cos\Bigl(g_\k\frac{\pi}{2}\Bigr) + \wt{B} \sin\Bigl(g_\k\frac{\pi}{2}\Bigr) = 0 \,,\quad
 \wt{A} \cos\Bigl(g_\k\frac{\pi}{2}\Bigr) - \wt{B} \sin\Bigl(g_\k\frac{\pi}{2}\Bigr) = 0 \,,
$$
from which we obtain two possibilities
\begin{eqnarray*}
 \wt{B} =0{\quad}{\rm and}{\quad} \cos\bigl(g_\k\frac{\pi}{2}\bigr)  &=& 0 \,,\cr
 \wt{A} =0{\quad}{\rm and}{\quad} \sin\bigl(g_\k\frac{\pi}{2}\bigr)  &=&  0\,.
\end{eqnarray*}
There are therefore two possible classes of solutions
\begin{enumerate}
\item[(a)] The coefficient $g_\k$ and quantum number $\mu$ are
given by
$$
  g_\k = g_{\k a} = 2 m + 1 \,,{\quad} \mu  = \mu_{a} = \k\,(2 m + 1)^2
$$
\item[(b)] The coefficient $g_\k$ and quantum number $\mu$ are
given by
$$
  g_\k = g_{\k b} = 2 m \,,{\quad} \mu  = \mu_{b} = \k\,(2 m)^2
$$
\end{enumerate}
In the case (a) the wave functions are given by
$$
Z_{ma}(z_x) =  \wt{A} \cos\Bigl((2 m + 1) \arcsin(\sqrt{\k}\,z_x)\Bigr)   \,.
$$
In the case (b) we obtain
$$
Z_{mb}(z_x) =  \wt{B} \sin\Bigl((2 m) \arcsin(\sqrt{\k}\,z_x)\Bigr)  \,.
$$

\begin{proposicion}
The eigenfunctions $Z_{ma}(z_x)$ and $Z_{mb}(z_x)$ of the problem
(\ref{EcSLZ(z)e}) are orthogonal in the interval  $[-\,a_\k,a_\k]$
with respect to the function $r= \mu/(1 - \k\,z_x^2)$.
\end{proposicion}
{\it Proof:} This statement is just a consequence of the properties 
of the Sturm-Liouville problems.

\subsection{Sturm-Liouville problem for the $Y$-equation}

The $\k$-dependent differential equation
$$
 a_0  p'' + a_1 \,p' + a_2 p = 0  \,,
$$
with
$$
 a_0 = 1 - \k\,y^2 \,,{\quad}  a_1 = - 2 \k\, (1+g_\k) y   \,,{\quad}
 a_2 = \nu - \k g_\k \,,
$$
is not self-adjoint since $a'_0 \ne a_1$ but it can be reduced to
self-adjoint form by making use of  an appropriate integrating factor
in such a way that we arrive to 
\begin{equation}
  \frac{d}{dy}\Bigl[\,q(y,\k)\,\frac{dp}{dy}\,\Bigr] + r(y,\k)\,p = 0 \,.
 \label{EcSLp(y)e}
\end{equation}
with $q=q(y,\k)$ and $r=r(y,\k)$ given by 
$$
  q(y,\k)  =  (1 - \k\,y^2)^{1+g_\k}  \,,{\quad}
  r(y,\k)  = (\nu - \k g_\k) (1 - \k\,y^2)^{g_\k}  \,, 
$$
Thus, this self-adjoint equation together with appropriate
conditions for the behaviour of the solutions at the end points,
constitute a  Sturm-Liouville problem.

If $\k$ is positive the range of the  variable $y$ is
limited by the restriction $y^2<1/\k$. In this case the
problem,  defined in the bounded interval
$[-\,a_\k,a_\k]$ with $a_\k=1/\sqrt{\k}$, is singular because
the function $q(y,\k)$ vanishes in the two end points
$y_1=-\,a_\k$ and $y_2=a_\k$.

From a purely mathematical viewpoint the eigenfunctions must be
finite when $y\to\pm 1/\sqrt{\k}$ (a continuous function in a
closed interval is always bounded and integrable). In addition,
from a quantum viewpoint, the wave functions $Y(y)$ must vanish
when  $y\to\pm 1/\sqrt{\k}$. But this second stronger condition is
satisfied because of the factor $(1 - \k\,y^2)^{(1/2)g_\k}$ in the
expression of $Y(y)$.

The eigenvalues are the quantized values of the parameter $\nu$,
that is,
$$
 \nu_{na} = \k[n(n+1) + g_{\k a}(2 n + 1)] \,,  {\quad} g_{\k a} = 2 m+1
 \,,  {\quad}  n=0,1,2,\dots
$$
$$
 \nu_{nb} = \k[n(n+1) + g_{\k b}(2 n + 1)] \,,  {\quad} g_{\k b} = 2 m 
 \,,  {\quad}  n=0,1,2,\dots
$$
and the eigenfunctions the
associated polynomial solutions.

\begin{proposicion}
The eigenfunctions of the problem (\ref{EcSLp(y)e}) are orthogonal
in the interval  $[-\,a_\k,a_\k]$ with respect to the function $r=
(1 - \k\,y^2)^{g_\k}$.
\end{proposicion}
{\it Proof:} This statement is just a consequence of the
properties of the Sturm-Liouville problems.

Because of this the polynomial solutions ${\cal P}_{mn}$,
$n=0,1,2,\dots$ of the equation (\ref{EcSLp(y)e}) satisfy
$$
  \int_{-\,a_\k}^{a_\k} {\cal P}_{mn_1}(y,\k)\,{\cal P}_{mn_2}(y,\k)
  (1 - \k\,y^2)^{m}\, dy = 0
  \,,\quad n_1\,\ne\,n_2 \,,\quad \k>0\,.
$$
The first even ${\cal P}_{mj}(y)$,  $j=0,2,4$, and odd ${\cal
P}_{mj}(y)$, $j=1,3,5$, polynomials of this orthogonal family are
\begin {enumerate}
\item Even index (even power polynomials)
\begin{eqnarray*}
 {\cal P}_{m0} &=&  1 \,,\cr
 {\cal P}_{m2} &=&  1 - \k (3 + 2 m) y^2  \,,\cr
 {\cal P}_{m4} &=&  1 - 2 \k (5 + 2 m) y^2 +  (\frac{\k^2}{3}) (5 + 2 m) (7 + 2 m) y^4
\end{eqnarray*}
\item  Odd index  (odd power polynomials)
\begin{eqnarray*}
 {\cal P}_{m1} &=&  y  \,,\cr
 {\cal P}_{m3} &=&  y - (\frac{\k}{3})(5 + 2 m)  y^3  \,,\cr
 {\cal P}_{m5} &=&   y - (\frac{2\k}{3})(7 + 2 m)  y^3
 + (\frac{\k^2}{15})(7 + 2 m)  (9  + 2 m)  y^5   \,.
\end{eqnarray*}
\end{enumerate}

If we define the  $\k$-dependent  functions  ${\cal Y}_{mn}$ by
$$
 {\cal Y}_{mn}(y,\k) = {\cal P}_{mn}(y,\k)(1 - \k\,y^2)^{m/2}
 \,,\quad  n=0,1,2,\dots
$$
then the above statement admits the following alternative form:
{\sl The $\k$-dependent functions ${\cal Y}_{mn}(y,\k) ={\cal
Y}_n(y,\k,m)$, $n=0,1,2,\dots$  are orthogonal with respect to the
weight function  $\wt{r}=1$}:
$$
 \int_{-\,a_\k}^{a_\k} {\cal Y}_{mn_1}(y,\k)\,{\cal Y}_{mn_2}(y,\k) \,dy  = 0
 \,,\quad n_1\,\ne\,n_2 \,,\quad \k>0\,,
$$
(the same $m$ in the two factors). Note that the orthogonality of the functions ${\cal Y}_{mn}(y,\k)$
coincides with the orthogonality with respect to the $y$-dependent
second factor of the measure $d\mu_\k$ discussed in the first
paragraph of Sec. \ref{Sec. 5}.

\subsection{Final solution}

  The wave functions  of the $\k$-dependent free particle in the sphere
$S_{\k}^2$, when written as functions of $(z_x,y)$, $z_x=x/\sqrt{\,1 - \k\,y^2}$,  corresponding to the first form $\Psi(e_1,e_{2j})$,  are given (up to a multiplicative constant) by
$$
 \Psi_{mnr}(z_x,y) =  Z_{mr}(z_x)\,Y_{mn}(y,\k)
  \,,{\quad}  r=a,b,   {\quad} m,n=0,1,2,\dots
$$
with
\begin{eqnarray*}
Z_{ma}(z_x) &=&  \cos\Bigl((2 m+1) \arcsin(\sqrt{\k}\,z_x)\Bigr)
\,,{\quad}  m= 0,1,2,\dots   \cr
Z_{mb}(z_x) &=&  \sin\Bigl((2 m) \arcsin(\sqrt{\k}\,z_x)\Bigr)
\,,{\quad}  m= 0,1,2,\dots
\end{eqnarray*}
and
$$
 Y_{mn}(y,\k)  = {\cal P}_{mn}(y,\k)(1 - \k\,y^2)^{m/2}
 \,,\quad  m, n=0,1,2,\dots
$$
with energies given by
\begin{eqnarray*}
 e_{m,n,a} = \mu_{ma} + \nu_{mna} &=& \k\,(2 m +1)^2 +
\k[n(n+1) + (2m+1)(2 n + 1)]   \cr
&=& \k (2 m + n +1) (2 m + n + 2) = \k (N+1)(N+2) \cr
 e_{m,n,b} = \mu_{mb} + \nu_{mnb} &=& \k\,(2 m)^2 +
\k[n(n+1) + (2m)(2 n + 1)]  \cr
 &=& \k (2 m + n ) (2 m + n + 1) = \k  N(N+1)
\end{eqnarray*}
So the total energy
$$
  E_{m,n,r}= \Bigl(\frac{\hbar^2}{2m}\Bigr) e_{m,n,r}
  \,,{\quad} r=a,b,   {\quad} m,n=0,1,2,\dots
$$
is proportional to the curvature $\k$ and depends only of total
quantum number $N$ given by $N= 2 m + n$. 

To sum up, the quantum free particle on the sphere $S_{\k}^2$ is
endowed with an infinite sequence of discrete energy values that
can be considered as a consequence of the compact nature of the
space.  The energy levels are not  equally spaced and the gap 
$\Delta E$, between two consecutives levels, is proportional to $N$. 
We recall that an energy eigenvalue $E$ is said to be degenerate 
when two or more independent eigenfunctions correspond to it. 
We have obtained that the values of $E_{m,n}$ depend only
on $N$ so they are degenerate with respect $m$ and $n$. 
Next we present the wavefunctions corresponding to the three lowest 
values of the energy: 
\begin{itemize}
\item[(i)]  The  fundamental level, with energy given by $e=2\,\k$, is non-degenerate and represented by only one wavefunction
$$
\Psi_{00a}(z_x,y) =  Z_{0a}(z_x)\,Y_{00}(y,\k) = 
\cos\bigl(\arcsin(\sqrt{\k}\,z_x)\bigr) \,.
$$
\item[(ii)]  The second  level, with energy given by $e=6\,\k$, is represented by the following two wavefunctions 
\begin{eqnarray*}
\Psi_{01a}(z_x,y) &=&  Z_{0a}(z_x)\,Y_{01}(y,\k) = 
\cos\bigl(\arcsin(\sqrt{\k}\,z_x)\bigr)\,y \,. \cr
\Psi_{10b}(z_x,y) &=&  Z_{1b}(z_x)\,Y_{10}(y,\k) = 
\sin\bigl(2\arcsin(\sqrt{\k}\,z_x)\bigr)\,(1 - \k\,y^2)^{1/2}\,.
\end{eqnarray*}
\item[(iii)]   The third  level, with energy given by $e=12\,\k$, is represented by the following three wavefunctions  
\begin{eqnarray*}
\Psi_{02a}(z_x,y) &=&  Z_{0a}(z_x)\,Y_{02}(y,\k) = 
\cos\bigl(\arcsin(\sqrt{\k}\,z_x)\bigr)\,(1 - 3\k\,y^2) \,.\cr
\Psi_{10a}(z_x,y) &=&  Z_{1a}(z_x)\,Y_{10}(y,\k) = 
\cos\bigl(3\arcsin(\sqrt{\k}\,z_x)\bigr)\,(1 - \k\,y^2)^{1/2} \,.\cr
\Psi_{11b}(z_x,y) &=&  Z_{1b}(z_x)\,Y_{11}(y,\k) = 
\sin\bigl(2\arcsin(\sqrt{\k}\,z_x)\bigr)\,y\,(1 - \k\,y^2)^{1/2} \,. 
\end{eqnarray*}

\end{itemize}

When we consider the Euclidean limit then, as $\k\to 0$,  all
these normalizable wave functions $\Psi_{m,n}$ disappear and the
associated energies $E_{m,n}$ vanish. This means that the
Euclidean wave planes characterized by a  continuous  spectrum
cannot be obtained as the limit of the discrete normalizable
$\k>0$ spectrum. This situation can be considered as a
consequence of the boundary conditions; that is, the  $\k\to 0$
limit of the general solutions of the $\k$-dependent equations
(without introducing boundary conditions in the points $-\,a_\k$
and $a_\k$) are the  Euclidean wave planes but once  the boundary
conditions are introduced in the points $-\,a_\k$ and $a_\k$ then
the result is a discrete $\k>0$ spectrum that cannot related with
the $\k=0$ description.

Finally,  we mention that another approach for the free motion in the sphere $S^3$, in terms of spectrum generating algebras, has  been recently developped in \cite{GaNeNiPrSa11}. 

\section{Hyperbolic $\k<0$ case }

\subsection{Resolution of the $Z$-equation}

If we assume a negative value for the curvature then we have
$\k=-|\k|<0$ and the  equation for $Z$ becomes
 \begin{equation}
 \bigl(1 +|\k|\,z_x^2\bigr)\,Z''   + \bigl(|\k|\,z_x\bigr)\,Z'  + \mu \,Z = 0 \,.  \label{EcZh}\end{equation}
 Then assuming for $Z$ an expression of the form
$$
   Z =   e^{{\ii}u(z_x)} \,,
$$
with $u(z_x)$ a function to be determined, we obtain that the
general solution of (\ref{EcZh}) is given by
\begin{equation}
  Z  =   A\,e^{{\ii}u(z_x)} + B\,e^{-{\ii}u(z_x)}  \,,{\quad}
  u = \frac{\sqrt{\mu}}{{\sqrt{|\k|}}}\,\arcsinh(\sqrt{|\k|}\,z_x) \,,\quad \k=-|\k|<0 \,,
\label{Zuh}\end{equation} that is a well defined function for all
the values of $z_x$ (we recall that in this case there are not
restrictions for the domain of $z_x$). This solution satisfies the
appropriate Euclidean limit
$$
 \lim_{\k\to 0}  Z = A\,e^{{\ii}k_x x} + B\,e^{-{\ii}k_x x}  \,,{\quad}  k_x = \sqrt{\mu} \,,
$$
and therefore it can be considered as representing a
$\k$-dependent hyperbolic deformation of the Euclidean plane-wave
solution.

\subsection{Resolution of the $Y$-equation}

The equation for the function $Y$ takes now the form
\begin{equation}
   (1 + |\k|\,y^2)\,Y'' + \bigl( 2|\k|\,y\bigr)\,Y'   +  \mu|\k|\,
   \Bigl( \frac{y^2 }{1 + |\k|\,y^2} \Bigr)\,Y +  \nu\,Y = 0 \,,{\quad} \k=-|\k|<0\,.
 \label{EcY(y)h}
\end{equation}
In order to obtain an hypergeometric equation, similar to the
spherical Eq. (\ref{Ecp(y)e1}), we can consider the following
factorization
$$
 Y(y) = p(y)\,(1 + |\k|\,y^2)^{(1/2)g} \,,
$$
but now we arrive to the condition $g^2 |\k| + \mu=0$ which leads to
$$
 g = {\ii}g_\k  \,,{\quad}  g_\k=\sqrt{\frac{\mu}{|\k|}} \,.
$$
The consequence is that the new equation must be complex. In fact
if we assume the complex factorization
$$
 Y(y) =  {\cal P}(y)\,(1 + |\k|\,y^2)^{({\ii}/2)g_\k} \,,\quad 
 {\cal P} =  p_1(y) +  {\ii} p_2(y) \,, 
$$
then we obtain
\begin{equation}
  (1 +  |\k|\,y^2) {\cal P}_{yy}  + 2 |\k| \bigl( 1 +  {\ii} g_\k\bigr)y {\cal P}_{y}  +\bigl(\nu   + {\ii}  |\k| g_\k \bigr){\cal P}  = 0  \,, \label{EcP(y)h}
\end{equation}
that represents a complex hypergeometric equation. Alternatively
it can be written as a system of two coupled real equations.
\begin{eqnarray*}
 (1 +  |\k|\,y^2)  p_{1yy}'' + 2|\k| y \,p_{1y}' + \nu p_1 &=&
 |\k| g_\k(2 y p_{2y}'+p_{2})  \cr
  (1 +  |\k|\,y^2)  p_{2yy}'' + 2|\k| y \,p_{2y}' + \nu p_2 &=&
-\, |\k| g_\k(2 y p_{1y}'+p_{1})
\end{eqnarray*}

So in this case there are two possibilities: (i) to solve directly
the Eq. (\ref{EcY(y)h}) (power series solution) or (ii) to solve
the complex hypergeometric equation (\ref{EcP(y)h}) (both
equations satisfy correctly the Euclidean limit). In any case it
can be proved the nonexistence of polynomial solutions for real
values of the quantum number $\nu$. This means that the
eigenvalues $\nu$ can take any positive value and the spectrum for
the energy is continuous as in the Euclidean case.

We note that in this hypergeometric case the two
variables, $z_x$ and $y$, are defined in the whole real line and
both functions, $Z(z_x)$ and $Y(y)$, turn out to be
nonnormalizable functions. So, in a sense, this hyperbolic case
can be considered as more similar to the Euclidean one  that the
spherical $\k>0$ one.

Finally, let us mention the study by Balazs and Voros \cite{BaVo86} of quantum mechanics on the hyperbolic plane. It is concerned, for the most part,   with the study of chaos in compact mani\-folds with constant negative curvature which arise as quotients of the hyperbolic plane (called pseudosphere in \cite{BaVo86}) by suitable discrete groups of isometries. In particular,  they discuss a pseudosphere analogous of the standard Euclidean plane waves, precisely those eigenfunctions of the corresponding Laplace-Beltrami operator which separate in horospherical coordinates. These can be imagined as the (suitably rescaled) limits of pseudospherical circular waves when the sink or source point goes to infinity. The solutions of the Schrodinger equation we are discussing here are neither circular waves nor horospherical waves, because these allow separation of variables in a variant of parallel coordinates. As mentioned in the introduction, and remarked also in \cite{BaVo86} the idea of `plane waves in a manifold of constant curvature' admits several possible realizations in spaces of constant negative curvature, and while the horospherical waves are somehow more natural than others (because there is still a source or sink at some point, albeit at infinity), the solutions we are dealing with here are a different possibility and would have a family of geodesics orthogonal to  given geodesic as wavefronts. In the $\k\to 0$ limit, these (as well as horospherical plane waves) can be expected to collapse to Euclidean plane waves. A more detailed study of solutions of the curved Laplace-Beltrami equations in different coordinates will be done elsewhere.

\section{$\k$-dependent Schr\"odinger equation  II}

The second alternative way of solving the quantum $\k$-dependent
problem is to consider the system of the two following Schr\"odinger equations
$$
\bigl(\widehat{H_1} + \k\, \widehat{J}^2\bigr)\,\Psi =
e_{1j}\,\Psi  \,,{\hskip 10pt} \widehat{H_2}\,\Psi = e_2\,\Psi \,,  
$$
that can be solved by using the property of separability of the
equation $ \widehat{H}\,\Psi = E\,\Psi $ in coordinates $(x,z_y)$
with $z_y$ defined as $z_y = y/\sqrt{1 - \k\,x^2}$\,.  

 Thus, if we assume that $\Psi(x,z_y)$ is a function of the form
$$
 \Psi(x,z_y)  = X(x) Z(z_y) \,,
$$
then we arrive to
$$
 \bigl(1 - \k\,z_y^2\bigr)\,Z''   - \bigl(\k\,z_y\bigr)\,Z'  + \mu' \,Z = 0 \,,$$
and
$$
   (1 - \k\,x^2)\,X'' - \bigl( 2\k\,x\bigr)\,X'   -  \mu'\k\,
   \Bigl(\frac{x^2}{1 - \k\,x^2} \Bigr)\,X +  \nu'\,X = 0   \,.
$$
where the two eigenvalues $\mu'$ and $\nu'$ are related with the two partial energies $e_2$ and $ e_{1j}$  by 
$$  
 \mu' = \Bigl(\frac{2 m }{\hbar^2}\Bigr) e_2 \,,{\quad}
 \nu' = \Bigl(\frac{2 m }{\hbar^2}\Bigr) e_{1j}\,.
$$
These two equations can be solved by repeating the previous analysis with the appropriate interchange of variables.  We only recall that this second approach leads to a value of $E$ given by $E = e_{1j} + e_2$ and that the solution $\Psi(x,z_y)$ can also be denoted by $\Psi(e_{1j}, e_2)$.

\section{Final comments and outlook}

Let us summarize our results. We have studied the quantum  free
particle on spherical and hyperbolic spaces using a curvature
dependent approach. In the first part of the paper, that was mainly
concerned with geometrical questions, an important point was the
identification of the three Killing vectors and the associated
Noether symmetries. This was important for the quantization
procedure that was carried out in two steps:
\begin{enumerate}
 \item[(i)]  Quantization of the three Noether momenta as self-adjoint
 operators with respect to a $\k$-dependent measure (of course, when $\k=0$
we recover the standard quantization of the linear and the angular
momenta).
 \item[(ii)] Construction of the quantum Hamiltonian $\widehat{H}(\k)$ as
a function of the three operators $\widehat{P_1}$,
$\widehat{P_2}$, and  $\widehat{J}$.
\end{enumerate}
The second part of the paper was devoted to the resolution of the
$\k$-dependent equations. The separation of the Schr\"odinger
equation in coordinates $(z_x,y)$ introduces the term depending of
the angular momentum $\widehat{J}$, that plays the role of an
effective potential,  in the the $y$-equation in such a way that
\begin{enumerate}
 \item[(i)]  The motion along the $z_x$ direction is a ($\k$-dependent) free
  motion and the solution is a $\k$-deformed plane wave.
 \item[(ii)] The motion along the $y$ direction leads to an hypergeometric
equation.
\end{enumerate}
What introduce differences between the $\k>0$ and the $\k<0$ cases
is that in the spherical case the space is compact and this leads
to a Sturm-Liouville problem with boundary conditions rather
similar to to the case of a quantum particle in a one-dimensional
square well with perfectly rigid impenetrable walls. The result is
a discrete spectrum with normalizable wave functions $\Psi_{m,n}$
and associated energies $E_{m,n}$ when $\k>0$.

We finalize pointing out two questions to be studied. First, as was stated 
in Sec. (\ref{Sec2}) this problem can also be solved by using $\k$-dependent spherical coordinates that corresponds to the approach $\bigl(\widehat{H_1} +
\widehat{H_2}\bigr)\,\Psi = e_{12}\,\Psi$, $\widehat{J}^2\,\Psi =
e_j\,\Psi$ (the solutions must be $\k$-deformations of the standard Euclidean spherical waves). Second, we have obtained, as a mathematical by-product
of this formalism, a $\k$-dependent family of orthogonal
polynomials. They deserve a deeper mathematical study.

\section{Appendix. Geodesic parallel coordinates}

Suppose $M$ be a 2-dimensional Riemannian manifold, $O$ a point on
$M$ and $g_1$ and $g_2$, two orthogonal geodesics through $O$. Let $P$ be
an arbitrary point, in some suitable neighbourhood of $O$, and
denote by $P_1$ and $P_2$ the orthogonal projections of $P$ on
$g_1$ and $g_2$ (that is, $P_1$ is the intersection of $g_1$ with the geodesic through $P$ orthogonal to $g_1$). Then we can characterize the point $P$ by
\begin{enumerate}
\item The two distances $(u,y_\k)$ defined as follows: $u$ is the distance of $O$ to $P_1$ (measured along $g_1$) and $y_\k$ the distance of $P_1$ to $P$ (measured along the geodesic by $P$ and $P_1$).

\item The two distances $(x_\k,v)$ defined as follows:   $x_\k$ is the distance of $P_2$ to $P$ (measured along the geodesic by $P$ and $P_2$) and $v$ the distance of $O$ to $P_2$ (measured along $g_2$).
\end{enumerate}
In the first case we have the parallel coordinates of $P$ relative
to $(O,g_1)$ and in the second case relative to $(O,g_2)$  \cite{Kl78}.
In the $(u,y_\k)$ system the curves `$u={\rm constant}$' are geodesics
and the curves `$y_\k=$constant' meet these geodesics orthogonally.
In the $(x_\k,v)$ system the geodesics are the curves `$v={\rm constant}$' 
and `$x_\k=$constant'.
Notice that in the general case we have $u{\ne}x_\k$ and $v{\ne}y_\k$.

 In the case of $M$ being a space of constant curvature $\k$, the $(u,y_\k)$
and $(x_\k,v)$ expressions for the differential arc length element $ds_\k^2$
are given by
\begin{equation}
  ds_{\k}^2 = \Cos_\k^2(y_\k)\,du^2 + dy_\k^2 \,,{\quad}{\rm and}{\quad}
  ds_{\k}^2 = dx_\k^2 + \Cos_\k^2(x_\k)\,dv^2 \,,
\end{equation}
so that in both cases we get $ds^2 = ds_{0}^2=dx^2 + dy^2$ for the
particular value $\k=0$ characterizing the Euclidean case.
These two systems, although different for $\k\ne 0$, can be
related by using formulae of spherical and hyperbolic trigonometry
for $\k>0$ and for $\k<0$ respectively.

\section*{Acknowledgments}

JFC and MFR acknowledges support from research projects
MTM-2009-11154 (MCI, Spain), and DGA-E24/1 (DGA, Spain);  
and MS from research projects MTM-2009-10751 (MCI, Spain) 
and JCyL-GR224-08 (JCyL, Spain).

{\small

    }
\vfill\eject

\end{document}